\documentstyle[12pt,aasms4]{article}

\lefthead{Weintraub et al.}
\righthead{NICMOS Observations of TWA Stars}

\begin{document}

\def\arcsec{{$^{\prime\prime}$}}
\def\ptsec{$^{\prime\prime}\mskip-7.6mu.\,$}
\def\ptdeg{$^\circ\mskip-7.6mu.\,$}
\def\Teff{T_{\rm eff}}
\def\wig#1{\mathrel{\hbox{\hbox to 0pt{%
          \lower.5ex\hbox{$\sim$}\hss}\raise.4ex\hbox{$#1$}}}}

\def\plottwo#1#2{\centering \leavevmode
\epsfxsize=.5\textwidth \epsfbox[100 150 750 650]{#1} \hfil
\epsfxsize=.5\textwidth \epsfbox[100 150 750 650]{#2}}


\def\captionbaselineskip{\baselineskip 20pt}
\def\textbaselineskip{\baselineskip 20pt}

\title{NICMOS Narrow-band Infrared Photometry of TW Hya Association Stars}

\author{David A. Weintraub\altaffilmark{1}, Didier Saumon\altaffilmark{1},
        Joel H. Kastner\altaffilmark{2},  
        and Thierry Forveille\altaffilmark{3}}

\altaffiltext{1}{Department of Physics \& Astronomy,
Vanderbilt University, P.O. Box 1807 Station B, Nashville, TN 37235}

\altaffiltext{2}{Carlson Center for Imaging Science, RIT, 
       84 Lomb Memorial Drive, Rochester, NY 14623}

\altaffiltext{3}{Observatoire de Grenoble, B.P. 53X, 38041 Grenoble Cedex, France}


\begin{abstract}

We have obtained 1.64, 1.90 and 2.15 $\mu$m narrow-band images of five 
T Tauri stars in the TW Hya Association (TWA) using the Near-Infrared 
Camera and Multiobject Spectrometer aboard the {\it Hubble Space 
Telescope}.   Most of the T Tauri stars in our study show evidence 
of absorption by H$_2$O vapor in their atmospheres; in addition, 
the low-mass brown dwarf candidate, TWA 5B, is
brighter at 1.9 $\mu$m than predicted by cool star models  
that include the effects of H$_2$O vapor but neglect dust.  We conclude
that the effect of atmospheric dust on the opacity is 
important at 1.9 $\mu$m for TWA 5B, the 
coolest object in our sample.  The available evidence suggests that the 
TWA is 5--15 MY old.  Comparison of the colors 
of TWA 5B with theoretical magnitudes as a function of age 
and mass then confirms previous claims that TWA 5B is substellar with a mass 
in the range 0.02--0.03 $\,M_\odot$.  The accurate single-epoch astrometry 
of the relative positions and separation of TWA 5A and  
TWA 5B reported here should permit the direct measurement of the orbital 
motion of TWA 5B within only a few years.  

\end{abstract}
\keywords{open clusters and associations: individual (TW Hydrae, 
CD$-$33$^\circ$7795)
--- stars: low-mass, brown dwarfs 
--- stars: pre-main sequence}

\section{Introduction}

For more than a decade, the young star TW Hya has been an enigma
since it lies in a region of sky apparently devoid of the raw materials to
form stars, nearly 13$^\circ$ from the nearest dark cloud, yet it is
unambiguously a classical T Tauri star (\cite{ruci1983}) 
surrounded by a great deal of cold  dust (\cite{wein1989}) and 
gas (\cite{zuck1995,kast1997}).
Recently, the TW Hya mystery was solved:  TW Hya, along with
other T Tauri stars found in an area of $\sim$100 square degrees 
of the southern sky (\cite{dela1989,greg1992}),
compose a uniquely close association of young stars known as the TW Hya 
Association (\cite{kast1997}).  

At a mean distance of only $\sim$55 pc, the TW Hya Association (hereafter 
TWA) is almost three times closer than the next nearest known region 
of recent star formation. Given the likely age ($\sim$10 MY) of the TWA,  
these stars could harbor very young planetary systems with fully formed
giant planets or low mass, brown dwarf companions, and may still be 
surrounding by circumstellar disks.  In fact, there is substantial 
evidence for circumstellar gas and dust around several of these 
stars (\cite{wein1989,zuck1993,zuck1995,kast1997}).
The relative proximity and the absence of significant interstellar or 
intra-molecular cloud extinction in the direction of the TWA 
make the prospects for detecting substellar 
companions around these nearest T Tauri stars much better than the 
prospects for similar searches for young low mass companions 
around T Tauri stars in Taurus-Auriga, Chamaeleon, Lupus or Ophiuchus, the
next closest regions of star formation. 
In a recent study of the TWA, Webb et al.~(1999) 
identified a total of at least 17 sources as members of the TWA.  In addition,
Lowrance et al.~(1999) and Webb et al.~have reported the 
discovery of a likely low mass brown dwarf companion (M $\simeq$ 0.02 
$\,M_\odot$) to TWA 5A (= CD$-$33$^\circ$7795) in a combination of 
ground-based and {\it Hubble Space Telescope} ({\it HST}) observations. 
TWA 5B is found almost 2\arcsec\ from TWA 5A; thus, despite its relative 
physical proximity ($\sim$100 AU) to the primary, TWA 5B is amenable to 
spectroscopic and astrometric studies, uncontaminated by light from TWA 5A.

In this paper, we report results from imaging the fields 
around five stars in the TWA, including the TWA 5 system, 
using the Near Infrared Camera and Multi-Object 
Spectrometer (NICMOS) and the {\it HST}.  The goal of this 
program was to search for companions  
around these stars.  Our choice of three narrowband filters centered 
at 1.64, 1.90 and 2.15 $\mu$m was designed to enable us to identify 
cool and low surface gravity objects, including substellar mass 
companions, through their likely strong signatures of H$_2$O absorption 
at 1.9 $\mu$m.

\section{Observations}

We obtained images of five star systems (Table 1) in the TWA 
using camera 1 (NIC1) and camera 2 (NIC2) of NICMOS between 
1998 May 30 and July 12 (U.T.)  
Observations of each of the five stars were made identically.  
Using NIC1 and filter F164N, we imaged 
each target in a four position, spiral dither pattern, 
with an integration time per position of 33.894 s.  Three images
were obtained at each position for 
a total integration time of 406.73 s.  We carried out identical
observations using NIC1 and filter F190N, with an integration 
time per position of 43.864 s and a total integration time of 
526.37 s.  Switching to NIC2 and filter F215N, we again obtained 
three sets of four-position dithered image suites; however, for 
the F215N observations we changed the starting position for the
dithered image suites in order to obtain a better median filtered
image for subtraction of the thermal background. The integration
time per position was 15.948 s for the NIC2 images, for a total 
integration time of 191.38 s.  

\section{Results}
\subsection{Imaging}

We find no sources in any of our images other than the previously 
known five primaries and four secondaries, to limiting magnitudes 
of 18.3, 18.4 and 17.5 in the F164N, F190N and F215N images, 
respectively, at distances beyond $\sim$1\ptsec3 at 1.64 and 1.90
$\mu$m and 2\ptsec3 at 2.15 $\mu$m. The images of TWA 5 (Fig.~1) 
reveal how how easy it is to detect and image young, intermediate mass, 
brown dwarf companions around stars in the TWA, even without a
coronagraph.
In addition, all nine imaged objects appear as point 
sources (with FWHM of 0\ptsec14, 0\ptsec16, and 0\ptsec18\ in the F164N,
F190N and F215N images, respectively), 
with no evidence (after deconvolutions performed with point spread functions 
[PSFs] generated using the software package Tiny 
Tim\footnote{http://scivax.stsci.edu/~krist/tinytim.html},  
direct subtractions of PSFs, and examinations of azimuthally averaged 
radial intensity profiles [Fig.~2])
of extended emission around any of them.  
Thus, although some of these stars appear to be 
surrounded by circumstellar material (e.g., TWA 1 is surrounded
by a circumstellar disk of radius $\sim$3\arcsec\ that is viewed nearly
face-on; \cite{wein1999,kris1999}), we conclude that these direct, narrow 
band images are insufficiently sensitive to image circumstellar disks 
around these stars.

\subsection{Photometry}

We report our photometry for these observations in Table~1.  
The factors used to convert from NICMOS count rates to absolute fluxes and 
magnitudes\footnote{http://www.stsci.edu/ftp/instrument\_news/NICMOS/NICMOS\_phot/keywords.html, 
version 1998, December 1} were 5.376665 $\times$ 10$^{-5}$ 
Jy sec ADU$^{-1}$ for the F164N filter, 4.866353 $\times$ 10$^{-5}$
Jy sec ADU$^{-1}$ for the F190N filter, and 3.974405 $\times$ 10$^{-5}$
Jy sec ADU$^{-1}$ for the F215N filter with zero point flux densities
of 1033 Jy, 862 Jy, and 690 Jy, respectively.
Photometry was obtained by measuring the total counts within a 0\ptsec5
radius aperture and then applying a correction factor of 1.15 to
compensate for the flux that falls outside of this radius\footnote{
http://www.stsci.edu/ftp/instrument\_news/NICMOS/nicmos\_doc\_phot.html}.

Figure 3 shows how the photometry of TWA stars through the  $J$, $H$, and $K$
broadband and the F164N, F190N and F215N narrowband filters  relates to the
near infrared spectral characteristics of late-type stars.  Stars with lower
effective temperatures have increasingly strong water absorption bands centered 
at $1.4\,\mu$m and $1.9\,\mu$m which are very effectively probed by this 
combination of broad and narrowband filters.

We find no systematic differences between the $K$ and F215N 
photometry (see Fig.~3, right panel), despite the slight difference 
in central wavelengths and large difference in bandwidth.  On the 
other hand the stars are systematically brighter at F164N than 
at $H$-band, the most extreme case being TWA 5B.  Finally,   
TWA 5B is the only star with an absolute flux that is clearly
lower at 1.9 than at 2.15 $\mu$m.

The left hand panel of Fig.~3 clearly shows how water absorption 
in the 1.35-1.55 and 1.7 - 2.1 $\mu$m regions will strongly affect broad 
band $H$ measurements but will have no effect on observations with the 
F164N filter (see also the library of near-infrared spectra published 
by Lan\c con \& Rocca-Volmerange 1992). Thus, 
the $H$-band and F164N observations reveal the presence of different
amounts of water vapor absorption in the spectra of most of the stars
in our sample.

\subsection{Astrometry}

For the four binary systems in our sample, we measured the intensity
centroids for each binary component and transformed the cartesian 
positions on the array into offsets in Right Ascension and Declination 
of the secondaries from the primaries (Table~2). 
Except for TWA 8, for which the binary separation is such that the
companion only appeared in the NIC2 images, 
the results presented in Table~2 are those obtained using only the 
NIC1 images since the spatial 
resolution is highest when using the NIC1 array.
The NIC1-based results in Table~2 are the statistical average and
standard deviations based on measurements of the F164N and F190N 
images.  The pixel to RA and 
Dec conversions were done using the plate scale measurement 
ephemeris generated by the NICMOS instrument team\footnote{
http://www.stsci.edu/ftp/instrument\_news/NICMOS/nicmos\_doc\_platescale.html}.
Comparison of the 
results for the F164N and F190N images indicate that, in most cases, 
we can determine image separations to an accuracy of 0.02 pixels ($<$ 1 milli-arcsec).

Because our astrometric results are obtained from unocculted {\it HST}
images and with the highest resolution camera in NICMOS, these results 
are much more precise than offsets previously reported for these 
binaries.  They are, however, consistent with previous results (Table~2).
In the case of TWA 5B, Lowrance, Weinberger \& Schneider (1999) 
recently independently determined that the offsets reported in 
Lowrance et al.~(1999) and Webb et al.~(1999) have a 
sign error in RA; the corrected values are reported in Table~2.

\section{TWA 5B} 

\subsection{The age of the TWA}

We have constructed an H-R diagram for the TWA (Fig.~4) using the
pre-main sequence tracks of Baraffe et al.~(1998).  This H-R diagram
is quite similar to that presented by Webb et al.~(1999), which is
based on the pre-main sequence tracks of D'Antona \& Mazzitelli (1997);
however, Fig.~4 appears to constrain the cluster age more tightly 
than does previous work on the TWA, presumably because of  
improved physics included in the Baraffe et al.\ tracks (see Baraffe 
et al.\ 1997 for a discussion).  Specifically, virtually all of the
stars, including TWA 5B, fall between the 3 and 10 MY isochrones. 
In comparison, the H-R diagrams of Webb et al. and Lowrance et 
al.~(1999) indicate that the TWA stars have ages in the range 1--100 MY
while Kastner et al.~(1997) suggested that 
the likely age of the TWA stars is 10--30 MY, based on lithium studies
(upper limit) and X-ray luminosities (lower limit).
TWA 6 and TWA 9A, which lie together almost on the
30 MY isochrone, and TWA 9B, which falls near the 100 MY isochrone,
are mild outliers in our and the Webb et al.
HR diagrams and appear to be older than the other TWA stars (however,
see Webb et al.~for other possible explanations).

What other information do we have to constrain the ages of the TWA stars?
Soderblom et al.~(1998) used the lithium abundance to place an age
range of 5--20 MY and a most probable age of 10 $\pm$ 3 MY 
on TWA 4 (HD 98800; EW(Li $\lambda$6708) = 0.36 \AA);  
Stauffer, Hartmann, \& Barrado y Navascues (1995) 
used the strength of the Li line to assign an upper limit of 
9--11 MY to TWA 11B (HR 4796B) while Jayawardhana et al.~(1998)
assigned an isochronal age of 8 $\pm$ 3 MY to this star; and 
Webb et al.~(1999) measured similar  
Li EW strengths for 14 of the 17 stars identified as members of the TWA 
and, on this basis, suggested that they are all less than $\sim$10 MY.
The excellent agreement between the ages estimated from the Li EWs and 
those obtained from photometry and pre-main sequence evolutionary tracks 
suggests that the age of the TWA is well constrained to be in 
the range 5--15 MY.  

\subsection{Mass and evolutionary status of TWA 5A and 5B}

TWA 5A is a M1.5 star (Webb et al. 1999) with $\Teff=3700 \pm 150\,$K (Leggett et al. 1996).
The distance to the TWA 5 system is presently unknown but can be estimated as $55 \pm 9\,$pc from
the measured parallaxes of four members of the association (Webb et al. 1999).  The range of
distances is consistent with the approximate angular dimension of the association.  With $d=55 \pm 9\,$pc
and $K=6.8 \pm 0.1$, we find $M_K=3.10 \pm 0.41$.  By comparing these values of $\Teff$ and $M_K$
with the evolution sequences of Baraffe et al. (1998), we find $M=0.75 \pm 0.15\,M_\odot$ and an age
of 2.5 to 6 MY for TWA 5A, assuming it is a single, pre-main sequence star (Fig. 4).  On the other hand,
Webb et al. report that TWA 5A is suspected to be a spectroscopic binary.  If we assume that TWA 5A is
binary with equal mass components, the mass of each component decreases to $0.7 \pm 0.15\,M_\odot$ and
the age range becomes 6 to 18 MY.

Since the Baraffe et al. (1998) sequence does not extend to substellar masses,
we analyze the photometric measurements of TWA 5B with evolutionary models
computed by Saumon \& Burrows (unpublished).  These models use the same interior
physics as Saumon et al. (1996) and Burrows et al. (1997) with the distinction that the surface
boundary condition is provided by the ``NextGen'' sequence of atmosphere models computed by
Allard and Hauschildt for cool stars (Allard et al. 1996, Hauschildt, Allard \& Baron 1999).
The atmospheric structures provide a surface boundary condition for the interior models by giving
a relation between the interior entropy (where the convective zone becomes essentially adiabatic
at depth) and the surface parameters $S(\Teff,g)$.  This relation plays a central role in controlling
the evolution of fully convective stars.  Colors are computed from the synthetic spectra,
and are therefore fully consistent with the evolution calculation.
This evolution sequence was calculated for objects with solar compositions and
masses between 0.01 and 0.3$\,M_\odot$, and is very similar to that of Baraffe et al. (1998) since it
uses the same input physics (equation of state, atmosphere models, nuclear reaction screening 
factors, etc.).  A limitation of the ``NextGen'' atmospheres is that
they do not include dust opacity, which becomes significant for $\Teff \wig< 2600\,$K.

Figure 5 shows the evolution of the absolute magnitudes at $I$, $J$, $H$ and $K$ bands, from 1 to 100 MY,
based on the models of Saumon \& Burrows (unpublished).  Each curve shows the evolution for a fixed mass.
The two dashed
lines highlight the 0.02 and 0.03$\,M_\odot$ models.  The boxes show the photometric measurements for
TWA 5B (Webb et al. 1999, Lowrance et al. 1999), with the height of the box representing the 
$\pm 1 \sigma$ photometric error and the width showing the 5--15 MY
estimated age of the association.  The absolute magnitudes of TWA 5B
assume a distance of 55 pc and the $\pm 9\,$pc uncertainty is shown by the error bar in the upper
right corner.
All four bandpasses indicate that the mass of TWA 5B is between 0.02 and 0.03$\,M_\odot$ with an upper
limit of $\sim 0.06\,M_\odot$ if the TWA 5 system lies on the far side of the association and near the
upper limit of our age estimate.  Given this mass range and the estimated age of TWA 5B, the models
indicate that its surface gravity is $3.8 \wig< \log g ({\rm cm/s}^2) \wig< 4.0$.

Stellar and substellar objects with $M \ge 0.012\,M_\odot$
undergo a phase of nearly constant luminosity which corresponds to the fusion of their primordial 
deuterium content (D'Antona \& Mazzitelli 1985, Saumon et al. 1996). This phase lasts for 2 - 20 MY
and contraction --- with a consequent steady decrease in luminosity --- resumes once
the deuterium is exhausted.  Figure 5 shows that TWA 5B is almost certainly in the deuterium burning 
phase of its evolution.

\subsection{Colors of TWA 5B}

The $IJK$ colors of TWA 5B are consistent with its dM8.5--dM9 spectral classification
based on 0.65--0.75 $\mu$m spectra (\cite{webb1999,legg1998}), and thus a temperature
of $\Teff = 2600 \pm 150\,$K 
(\cite{luhm1997,legg1996}).  In a $J-H$ vs. $H-K$  
diagram, TWA 5B falls well outside of the observed sequence of very-low 
mass stars and brown dwarf candidates in the field (\cite{legg1998}),
while all other members of the association fall along the observed sequence
of field stars. This indicates that the $H$ magnitude for TWA 5B 
may be erroneous (by $\ge$ 1$\sigma$) or that its relatively low surface gravity 
results in a redder $H-K$ color. 

The narrowband infrared colors are shown in Fig.~6 along with the 
synthetic colors from the ``NextGen'' spectra.  Each curve shows the colors 
for $\Teff=2600$, 2800, 3000 and 3200$\,$K (from left to right) for a 
fixed gravity.  The colors of TWA 5B are shown by the triangle with 
error bars.  
For the estimated $\Teff = 2600 \pm 150\,$K and $\log g = 3.9 \pm 0.1$, 
there is a reasonable agreement for the F164N$-$F215N  color but the models 
are $\sim 0.4$ magnitude too blue in F164N$-$F190N.  Consequently,
TWA 5B is brighter at 1.9 $\mu$m than predicted by the models.

The F190N bandpass falls 
in the middle of a strong H$_2$O absorption band (Fig.~3) whose strength 
probably is overestimated by the ``NextGen'' models. 
Allard et al.~(1997) compare a sequence of near-infrared 
spectra of late M dwarfs with their synthetic spectra.  In all cases,
the models overestimate the depth of the H$_2$O band, an effect which 
increases for later spectral types.  While an inadequate H$_2$O opacity 
may be partly responsible for this effect, Tsuji, Ohnaka, \& Aoki 
(1996) have shown that the condensation of dust in atmospheres of low 
$\Teff$ results in a source of continuum opacity which decreases the depth 
of the water absorption bands. New atmosphere models including dust 
opacity (Tsuji, Ohnaka, \& Aoki; Leggett, Allard \& Hauschildt 1998) 
indicate that its effects on the spectrum (and on broadband colors) 
become discernible for $\Teff \wig< 2800\,$K but remain moderate 
($\sim 0.1$ mag) at the effective temperature of TWA 5B ($\sim 2600\,$K).  
While current models including dust opacity may not fully account for the relatively
high F190N flux of TWA 5B, the F164N$-$F190N color of TWA 5B is a strong 
indication of the presence of dust in its atmosphere.

\subsection{Astrometry of TWA 5B}

At a distance of 55 $\pm$ 9 pc, TWA 5B 
lies at a projected distance from TWA 5A of 108 astronomical 
units. TWA 5A  has a spectral type of M1.5 and a likely 
mass for the central binary of $\sim$1.4 M$_\odot$ while 
TWA 5B has an estimated 
spectral type of M8.5 (\cite{webb1999}) and a likely mass
of $\sim$25 M$_{jup}$. Given this information about the 
TWA 5 system, the orbital period P of TWA 5B 
should be P $\simeq$ 1000 years.  Thus, the angular motion of 
TWA 5B, assuming a circular orbit of radius 1\ptsec96 
viewed nearly pole-on, would be 0\ptsec013 yr$^{-1}$ 
(or 0\ptsec010 yr$^{-1}$ if TWA 5A is a single star with a mass of 
0.7 M$_\odot$ and P = 1300 years). 

The NICMOS observations of TWA 5 were obtained on 25 April 
(\cite{lowr1999a}) and 12 July 1998 (this paper), a difference of 
$\sim$0.21 years.  In only one-fifth of a 
year, the orbital motion of TWA 5B would have changed its
position relative to TWA 5A by only $\sim$0\ptsec0027, too small for the 
positional difference to be measured at these two epochs.
Thus, the differences between the positions we measured and the 
corrected positions from earlier epoch observations reported by 
Lowrance, Weinberger, \& Schneider (1999) are strictly
due to the relative accuracies of the different measurements.

Although we have demonstrated that the positional change for TWA 5B
is measurement error, not orbital motion, we also have shown that
the position reported in this paper is a very accurate ``starting'' 
position for TWA 5B.  In addition, our results show that it is possible
to measure the relative separation of these two objects to an accuracy of
only a few thousandths of an arcsec.  Thus, the orbital motion of TWA 
5B should be measurable to a fairly high degree of accuracy with 
ground-based observing facilities equipped with adaptive optics or 
with the refurbished NICMOS camera.

\section{Summary}

To the sensitivity limits of these data, our images 
reveal no detectable circumstellar disks or infrared reflection
nebulae, and no low mass stellar or substellar companions around stars 
in the five 
studied TWA systems other than the previously discovered TWA 5B.
As for TWA 5B itself, our results suggest that this object
has a mass in the range of 0.02--0.03 $\,M_\odot$, in good agreement with
the work of Lowrance et al.~(1999).  Finally, while our single epoch
observations cannot demonstrate or measure the orbital 
motion of TWA 5B, they are more than accurate
enough to permit the measurement of this motion, in combination
with future epoch {\it HST} or adaptive optics, ground-based 
observations, with a baseline of only about a year.  

\acknowledgments{We thank the referee for thoughtful suggestions 
which improved the clarity of the manuscript, 
A. Burrows for computing the evolutionary 
sequences used in this work, and F. Allard, P.H. Hauschildt, I. Baraffe 
and G. Chabrier for making their synthetic spectra and models available. 
This research was supported by NSF grant 
AST93-18970 and NASA grants NAG5-4988 and GO07861.01-96A and  
is based on observations obtained with the NASA/ESA {\it Hubble Space 
Telescope} at the Space Telescope Science Institute, which is operated by 
the Association of Universities for Research in Astronomy, Inc., under NASA 
contract NAS5-26555.}

\begin{deluxetable}{cllrrrrr}
\tablewidth{520pt}
\tablenum{1}
\tablecaption{Measured Photometry of TWA Stars}
\tablehead{
               \colhead{TWA}
             & \colhead{Common}
             & \colhead{Spec.}
             & \colhead{F164N}
             & \colhead{$H$}
             & \colhead{F190N}
             & \colhead{F215N}
             & \colhead{$K$}
\\
               \colhead{Number}  
             & \colhead{Name}
             & \colhead{Type\tablenotemark{a}}
             & \colhead{(mag)\tablenotemark{b}}
             & \colhead{(mag)\tablenotemark{c}}
             & \colhead{(mag)\tablenotemark{b}}
             & \colhead{(mag)\tablenotemark{b}}
             & \colhead{(mag)\tablenotemark{c}}
          }
\startdata
1  & TW Hya               & K7  &  7.50$\pm$0.01 &  7.65$\pm$0.1 &  7.53$\pm$0.01 &  7.41$\pm$0.01 &  7.37$\pm$0.07 \nl
2A & CD$-$29$^\circ$8887A & M0.5&  7.26$\pm$0.01 &  \nodata      &  7.39$\pm$0.01 &  7.20$\pm$0.01 &  7.18$\pm$0.07 \nl
2B & CD$-$29$^\circ$8887B & M2  &  8.08$\pm$0.01 &  \nodata      &  8.12$\pm$0.01 &  7.94$\pm$0.01 &  7.99$\pm$0.07 \nl
3A & Hen 600A             & M3  &  7.52$\pm$0.01 &  7.60$\pm$0.1 &  7.60$\pm$0.01 &  7.37$\pm$0.01 &  7.28$\pm$0.07 \nl
3B & Hen 600B             & M3.5&  7.90$\pm$0.01 &  8.07$\pm$0.1 &  7.98$\pm$0.01 &  7.79$\pm$0.01 &  7.80$\pm$0.07 \nl
5A & CD$-$33$^\circ$7795A & M1.5&  6.93$\pm$0.01 &  7.06$\pm$0.1 &  7.00$\pm$0.01 &  6.86$\pm$0.01 &  6.83$\pm$0.07 \nl
5B & CD$-$33$^\circ$7795B & M8.5& 11.76$\pm$0.01 & 12.1$\pm$0.1  & 11.83$\pm$0.01 & 11.49$\pm$0.01 &  11.5$\pm$0.07 \nl
8A & USNO 21A             & M2  &  7.59$\pm$0.01 &  7.72$\pm$0.1 &  7.66$\pm$0.01 &  7.53$\pm$0.01 &  7.44$\pm$0.07 \nl
8B & USNO 21B             & M5  & \nodata\tablenotemark{d} & \ 9.36$\pm$0.1 & \nodata\tablenotemark{d} &  \ 9.13$\pm$0.01 &  \ 9.01$\pm$0.07 \nl
\enddata
\tablenotetext{a}{from Webb et al.~(1999).}
\tablenotetext{b}{this paper.}
\tablenotetext{c}{from Webb et al.~(1999). Also, Lowrance, Weinberger, \& Schneider 1999
                  report $H$ = 7.2$\pm$0.1 
                  for TWA 5A and $H$ = 12.14$\pm$0.06 and $K$ = 11.4$\pm$0.2 for TWA 5B.}
\tablenotetext{d}{TWA 8A and 8B are more widely spaced than the size of the NIC1 field of view; hence, 
                  at F164N and F190N we were able to obtain data only for the 
                  primary.}
\end{deluxetable}

\clearpage


\begin{deluxetable}{cllllc}
\tablewidth{510pt}
\tablenum{2}
\tablecaption{Measured Binary Separations\tablenotemark{a}}
\tablehead{    \colhead{Source}
             & \colhead{$\Delta$RA}
             & \colhead{$\Delta$Dec}
             & \colhead{Separation}
             & \colhead{Position}
             & \colhead{Ref.\tablenotemark{b}}
\\
               \colhead{}  
             & \colhead{(arcsec)}
             & \colhead{(arcsec)}
             & \colhead{(arcsec)}
             & \colhead{Angle}
             & \colhead{}
          }
\startdata
TWA 2B\tablenotemark{c} &  $+$0.281$\pm$0.001  & $+$0.492$\pm$0.001 & 0.567$\pm$0.001  &  29.73$^\circ$$\pm$0.03$^\circ$ & 1\nl
                        &  $+$0.3\ \ \ $\pm$0.1& $+$0.5\ \ \ $\pm$0.1 & 0.6\ \ \ $\pm$0.1   & 31$^\circ$\ \ \ \ $\pm$3$^\circ$   & 2\nl 
TWA 3B\tablenotemark{d} &  $-$0.866$\pm$0.001  & $-$1.187$\pm$0.001 & 1.469$\pm$0.001  & 216.11$^\circ$$\pm$0.03$^\circ$ & 1\nl
                        &  $-$0.8\ \ \ $\pm$0.1\ \ \ & $-$1.2\ \ \ $\pm$0.1 & 1.4\ \ \ $\pm$0.1      & 214$^\circ$\ \ \ $\pm$3$^\circ$              & 2\nl
TWA 5B\tablenotemark{e} &  $-$0.038$\pm$0.001  & $+$1.960$\pm$0.006 & 1.960$\pm$0.006  &$-$1.11$^\circ$$\pm$0.03$^\circ$ & 1 \nl
                        &  $-$0.1\ \ \ $\pm$0.1\ \ \ & $+$1.9\ \ \ $\pm$0.1 & 1.9\ \ \ $\pm$0.1      & $-$3$^\circ$\ \ \ $\pm$3$^\circ$                & 3\nl
                        &  $-$0.04\ \ $\pm$0.01\ \ \ & $+$1.95\ $\pm$0.01 & 1.96\ $\pm$0.01    &$-$1.2$^\circ$\ \ $\pm$0.1$^\circ$  & 3\nl
TWA 8B\tablenotemark{f} &  $-$1.220$\pm$0.014   & $-$13.162$\pm$0.019 & 13.219$\pm$0.021  & 185.30$^\circ$$\pm$0.06$^\circ$ & 1 \nl
                        &  $-$1.3\ \ \ $\pm$0.1 & $-$13.0\ \ \ \ $\pm$0.1  & 13.0\ \ \ $\pm$0.1  & 186$^\circ$\ \ \ $\pm$3$^\circ$ & 2\nl
\tablebreak
\enddata
\tablenotetext{a}{from primary to secondary, based on centroid positions in NIC1 array images for 
 TWA 2B, TWA 3B and TWA 5B. As TWA 8B was only in the field of view in the NIC2 images, this offset 
 is taken from those NIC2 images.}
\tablenotetext{b}{ 1 = this paper. 2 = \cite{webb1999}. 3 = \cite{lowr1999b}.  
Lowrance, Weinberger \& Schneider report corrections of the position of TWA 5B
 originally reported in both Lowrance et al.~(1999) and Webb et al.~(1999).} 
\tablenotetext{c}{NIC1 plate scale: X=0.0431862 arcsec px$^{-1}$, Y=0.0430120 arcsec px$^{-1}$,
 based on plate scale measurements from June 4, 1998.  Observations obtained May 30, 1998.}
\tablenotetext{d}{NIC1 plate scale: X=0.0431887 arcsec px$^{-1}$, Y=0.0430144 arcsec px$^{-1}$,
 interpolated from plate scale measurements obtained on June 4 and August 6, 1998. Observations 
 obtained July 1, 1998.}
\tablenotetext{e}{NIC1 plate scale: X=0.0431897 arcsec px$^{-1}$, Y=0.0430154 arcsec px$^{-1}$,
 interpolated from plate scale measurements obtained on June 4 and August 6, 1998. Observations 
 obtained July 12, 1998.}
\tablenotetext{f}{NIC2 plate scale: X=0.0759831 arcsec px$^{-1}$, Y=0.0753005 arcsec px$^{-1}$,
 interpolated from plate scale measurements obtained on June 4 and August 6, 1998. Observations 
 obtained July 9, 1998.}
\end{deluxetable}

\clearpage
\subsection*{FIGURE CAPTIONS}

\figcaption{
NICMOS Images of TWA 5A/B, as seen through narrowband filters at 1.64 
$\mu$m (F164N) (1 pixel = 0\ptsec043), 1.90 $\mu$m (F190N) (1 pixel = 
0\ptsec043), and 2.15 $\mu$m (F215N) (1 pixel = 0\ptsec076). 
All the small scale features seen in the images except for TWA 5B 
are seen identically in the model point spread functions for NICMOS.
}

\figcaption{
Azimuthally averaged radial intensity profiles for five of the TWA stars
in our sample as seen in the a) F164N (1 pixel = 0\ptsec043), b) F190N 
(1 pixel = 0\ptsec043) and c) F215N  (1 pixel = 0\ptsec076) filters. 
Each profile is normalized to the brightness of the 
central pixel in that profile.  All profiles look identical to the 
PSFs generated by Tiny Tim, except for the presence of known companions.
}

\figcaption{
Photometry of TWA stars.
Left panel:  Synthetic spectra for a gravity of $\log g=4$ and 
$\Teff=4000\,$K ($\sim$K7) to 2600$\,$K ($\sim$M6), from top to bottom, respectively 
(Allard et al.~1996, Hauschildt et al.~1999).  Bars at the bottom show
the bandpasses of the Johnson-Cousin $J$, $H$ and $K$ filters and of the 
narrowband NICMOS filters F164N, F190N and F215N. Right panel: Broadband 
(Webb et al.~1999, Lowrance et al.~1999) and narrowband photometry for a 
representative sample of TWA stars.  The photometry is expressed in 
arbitrary flux units, normalized to the $J$ flux.  Boxes show the width of
the broadband filters and $\pm1\sigma$ error bars. Solid squares 
represent narrowband photometry results. The stars are ordered 
by spectral type, with later types at the bottom.
}

\figcaption{
Evolutionary tracks from Baraffe et al.~(1998) and the stars of 
the TW Hya association.  Solid lines show the evolution of stars of masses 
from 1.0 to 0.1$\,M_\odot$ in steps of 0.1$\,M_\odot$ and for 0.08$\,M_\odot$, 
from left to right, respectively. Isochrones for $\log t{\rm (yr)} =6.5$, 
7, 7.5, and 8 are shown by dotted curves.  The effective temperatures of the 
TWA stars were obtained from the Webb et al.~(1999) spectral types and the 
Luhman \& Rieke (1998) spectral type-$T_{\rm eff}$ relation.  The
error bars on $M_K$ reflect the uncertainties in the photometry 
(Webb et al.~1999) and on the distances.
}

\figcaption{
Evolution of $M_I$, $M_J$, $M_H$, and $M_K$ magnitudes for very
low mass stars and brown dwarfs.  From top to bottom, the curves correspond 
to masses of 0.2, 0.15, 0.125, 0.1, 0.09, 0.08, 0.07, 0.06, 0.05, 0.04, 0.03, 
0.02, and 0.01$\,M_\odot$, respectively.  Heavier curves show objects with
$M \ge 0.08\,M_\odot$ which eventually become main sequence stars.  
}

\figcaption{NICMOS narrow-band color-color diagram.  Colors calculated 
from the NextGen synthetic spectra (Hauschildt et al.~1999)
are shown by the dots.  Models with the 
same gravity are connected by a solid line, with $\log g\,({\rm cgs}) = 3.5$ 
to 5.5 from top to bottom, respectively.  The effective temperatures
shown are 2600, 2800, 3000, and 3200$\,$K, increasing from left to right.  
The observed colors of TWA5 B are shown by a triangle.
}


\begin{thebibliography}{}
\bibitem[Allard et al.\ 1997]{alla1997}
Allard, F., Hauschildt, P. H., Alexander, D. R., \& Starrfield, S. 1997,
ARAA, 35, 137

\bibitem[Allard et al.\ 1996]{alla1996}
Allard, F. Hauschildt, P. H., Baraffe, I. \& Chabrier, G. 1996, 
ApJ, 465, L123


\bibitem[Baraffe et al.\ 1997]{bara1997}
Baraffe, I., Chabrier, G., Allard, F., \& Hauschildt, P. H. 1997, A\&A, 327, 1054

\bibitem[Baraffe et al.\ 1998]{bara1998}
Baraffe, I., Chabrier, G., Allard, F., \& Hauschildt, P. H. 1998, A\&A, 337, 403

\bibitem[Burrows et al.\ 1997]{burr1997}
Burrows, A., Marley, M. S., Hubbard, W. B.,  Lunine, J. I., Guillot, T.,
Saumon, D., Freedman,  R. S., Sudarsky, D., \& Sharp, C. 1997, ApJ,
491, 856

\bibitem[D'Antona \& Mazzitelli 1985]{dant1985}
D'Antona, F., \& Mazzitelli, I. 1985, ApJ, 296, 502


\bibitem[D'Antona \& Mazzitelli 1997]{dant1997}
D'Antona, F., \& Mazzitelli, I. 1997, Mem. Soc. Astron. Italiana, 68, 4

\bibitem[de la Reza et al.\ 1989]{dela1989}
de la Reza, R., Torres, C. A. O., Quast, G., Castillo, B. V., 
\& Vieira, G. L. 1989, ApJ, 343, L61

\bibitem[Gregorio-Hetem et al.\ 1992]{greg1992}
Gregorio-Hetem, J., L\'epine, J. R. D., Quast, G., Torres, C. A. O., 
\& de la Reza, R. 1992, AJ, 103, 549

\bibitem[Hauschildt, Allard \& Baron 1999]{haus1999}
Hauschildt, P. H., Allard, F. \& Baron, E. 1999, ApJ, 512, 377

\bibitem[Jayawardhana et al.\ 1998]{jaya1998}
Jayawardhana, R., Fisher, S., Hartmann, L., Telesco, C., 
Pina, R. \& Fazio, G. 1998, ApJ, 503, L79

\bibitem[Kastner et al.\ 1997]{kast1997} 
Kastner, J. H., Zuckerman, B., Weintraub, D. A., \& Forveille, T. 1997,
Science, 277, 67 

\bibitem[Krist et al.\ 1999]{kris1999}
Krist, J. E., Stapelfeldt, K. R., Burrows, C. J., Menard, F., Padgett, D. L.
1999, BAAS, 31, 935

\bibitem[Lan\c con \& Rocca-Volmerange 1992]{lanc1992}
Lan\c con, A. \& Rocca-Volmerange, B. 1991, A\&ASS, 96, 593

\bibitem[Leggett et al.\ 1996]{legg1996}
Leggett, S. K., Allard, F., Berriman, G., Dahn, C. C. \& Hauschildt, P. H.
1996, ApJS, 104, 117
                 
\bibitem[Leggett, Allard \& Hauschildt 1998]{legg1998}
Leggett, S. K., Allard, F., \& Hauschildt, P. H. 1998, ApJ, 509, 836

\bibitem[Lowrance et al.\ 1999]{lowr1999a}
Lowrance, P. J. et al. 1999, ApJ, 512, L69 

\bibitem[Lowrance, Weinberger \& Schneider 1999]{lowr1999b}
Lowrance, P. J., Weinberger, A. \& Schneider, G., 1999, pers. comm. 

\bibitem[Luhman, Liebert \& Rieke 1997]{luhm1997}
Luhman, K. L., Liebert, J., \& Rieke, G. H. 1997, ApJ, 489, L165

\bibitem[Luhman, \& Rieke 1997]{luhm1998}
Luhman, K. L., \& Rieke, G. H. 1998, ApJ, 497, 354

\bibitem[Rucinski \& Krautter 1983]{ruci1983}
Rucinski, S. M., \& Krautter, J. 1983, A\&A, 121, 217

\bibitem[Saumon et al.\ 1996]{saum1996}
Saumon, D., Hubbard, W. B., Burrows, A., Guillot, T., Lunine, J. I., 
\& Chabrier, G. 1996, Apj, 460, 993

\bibitem[Soderblom et al.\ 1998]{sodo1998}
Soderblom, David. R. et al. 1998, ApJ, 498, 385.

\bibitem[Stauffer, Hartmann \& Barrado y Navascues 1995]{stau1995}
Stauffer, J. R., Hartmann, L. W., \& Barrado y Navascues, D. 1995, ApJ, 454, 910

\bibitem[Tsuji, Ohnaka \& Aoki 1996]{tsuj1996}
Tsuji, T., Ohnaka, K., \& Aoki, W. 1996, A\&A, 305, L1

\bibitem[Webb et al.\ 1999]{webb1999} 
Webb, R. A., Zuckerman, B., Patience, J., White, R. J., Schwartz, M. J.,
McCarthy, C., \& Platais, I. 1999, ApJ, 512, L63

\bibitem[Weintraub, Sandell \& Duncan 1989]{wein1989}
Weintraub, D. A., Sandell, G., \& Duncan, W. D. 1989, ApJ, 340, L69

\bibitem[Weinberger et al.\ 1999]{wein1999}
Weinberger, A. J., Schneider, G., Becklin, E. E., Smith, B. A., Hines, D. C.
1999, BAAS, 31, 934

\bibitem[Zuckerman \& Becklin 1993]{zuck1993} 
Zuckerman, B. \& Becklin, E. E. 1993, ApJ, 406, L25

\bibitem[Zuckerman et al. 1995]{zuck1995} 
Zuckerman, B., Forveille, T., \& Kastner, J.H. 1995, Nature, 373, 494
\end{thebibliography}
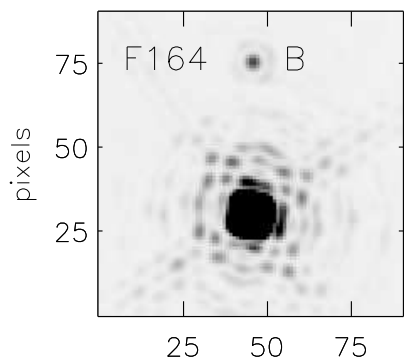
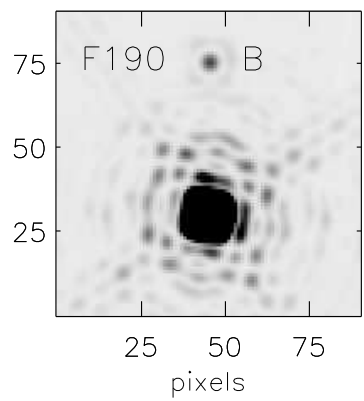
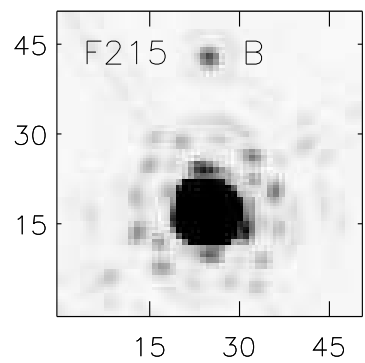
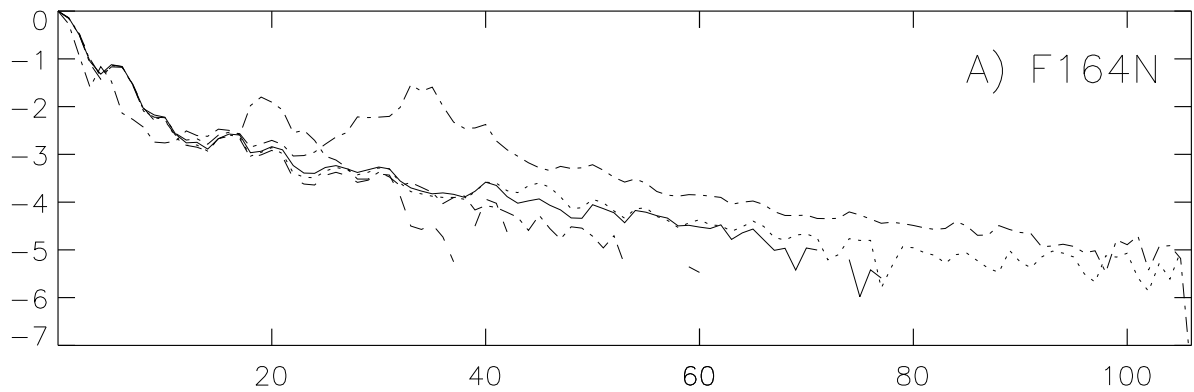
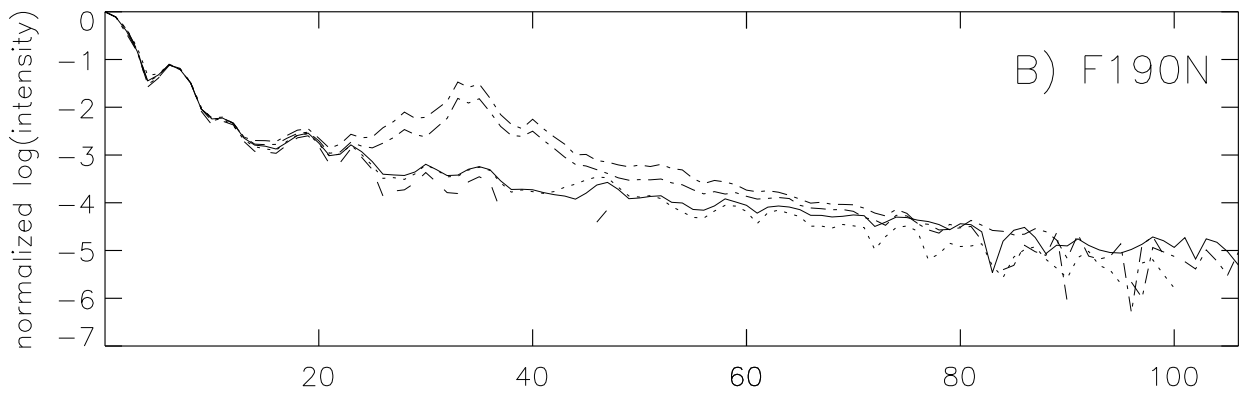
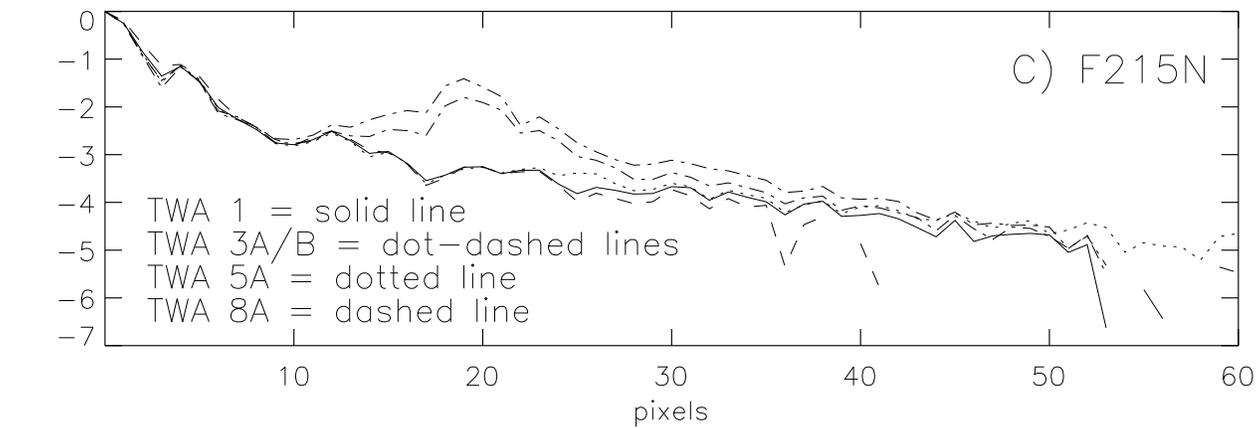
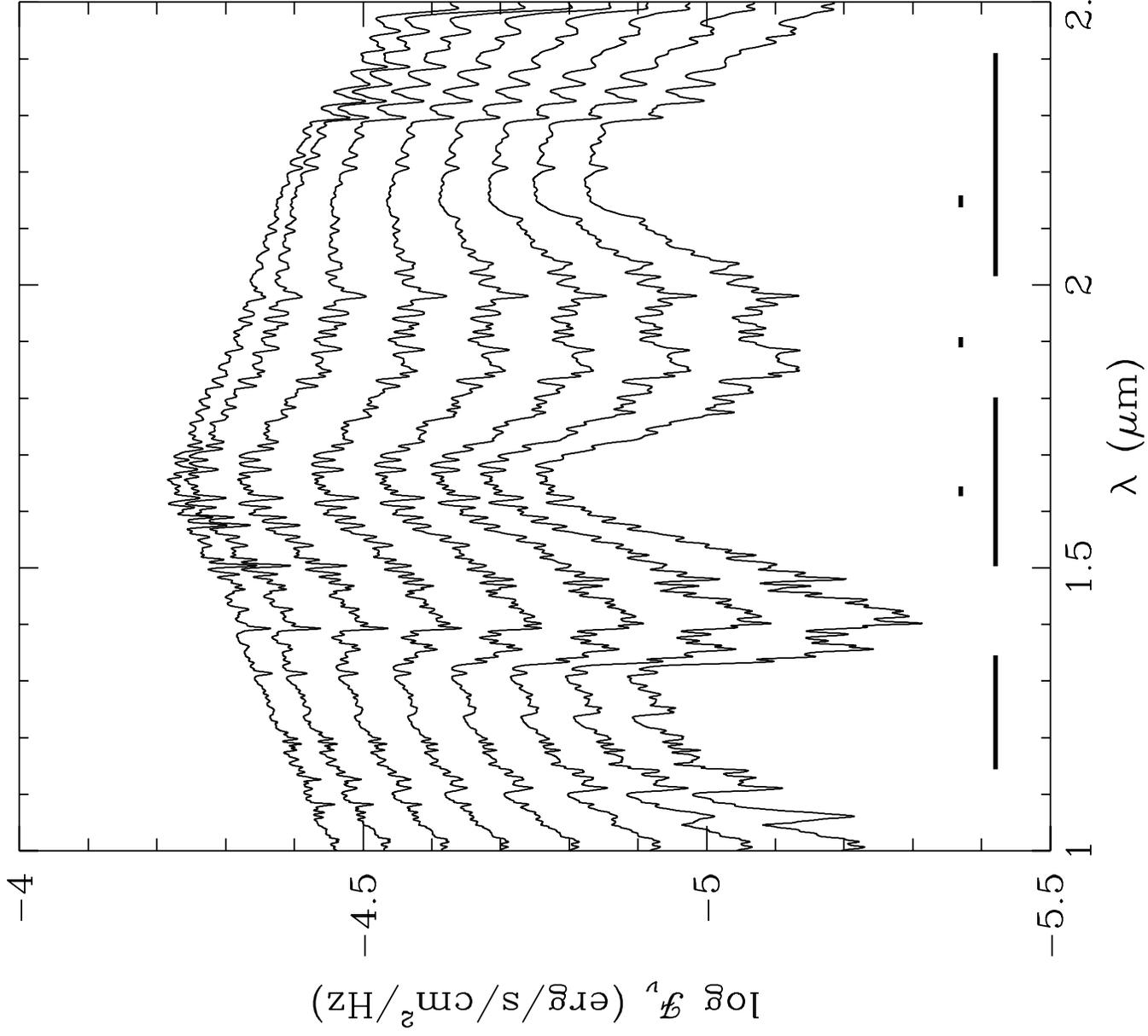
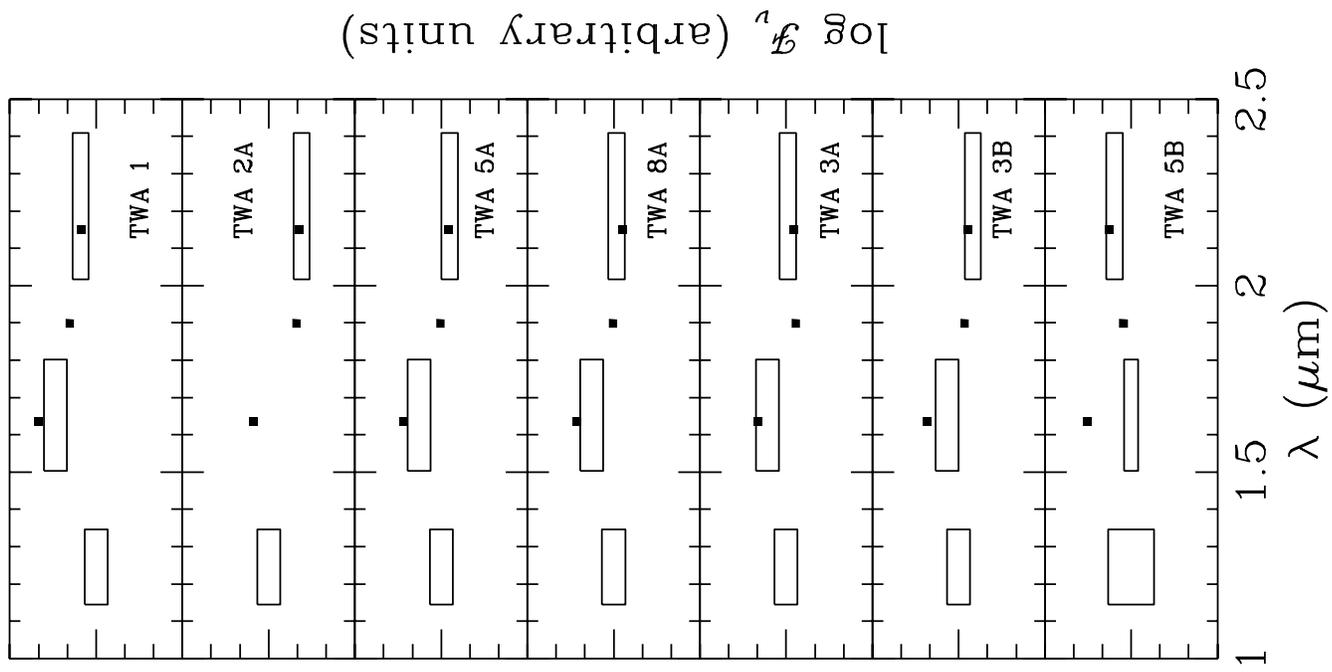
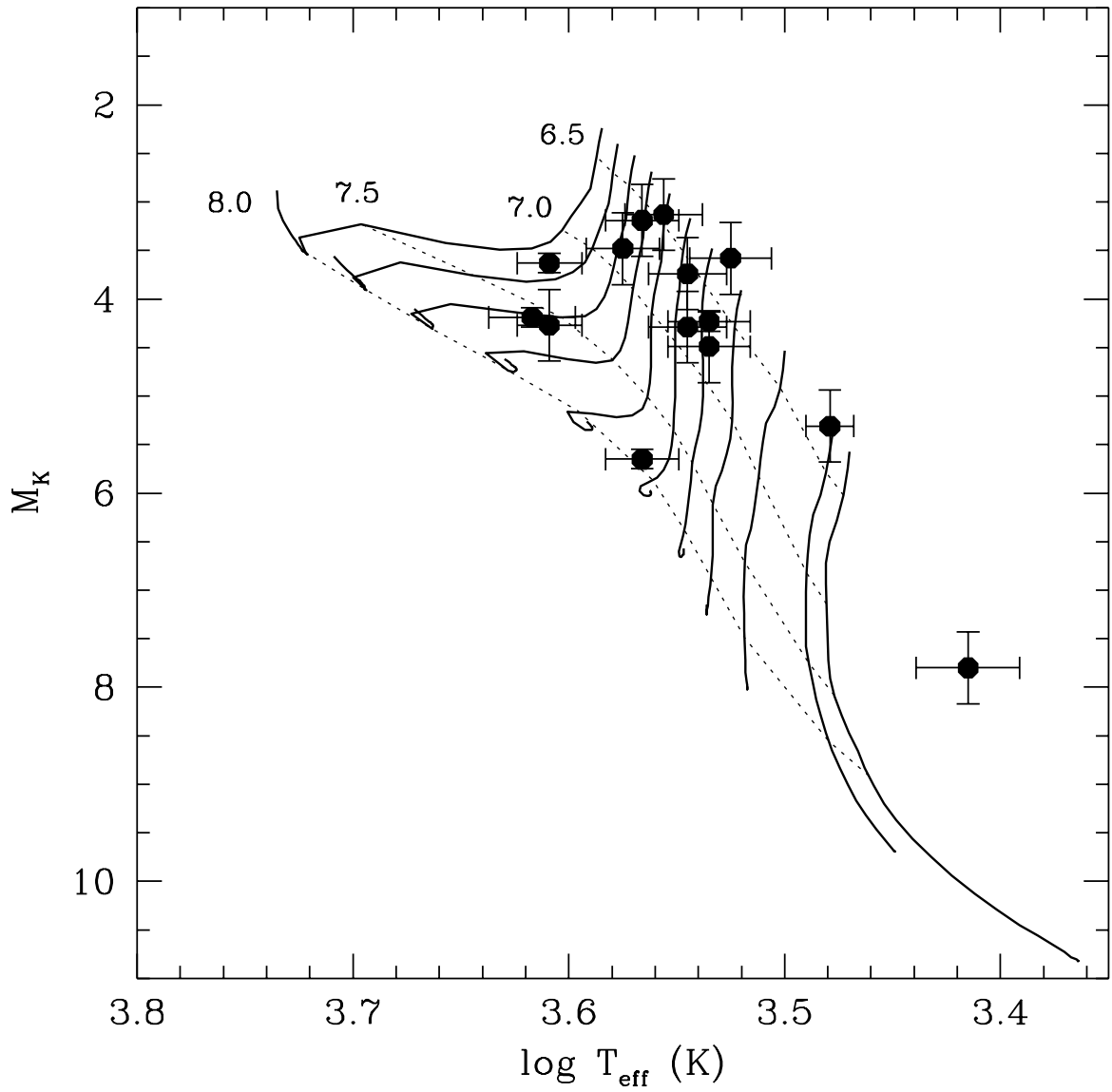
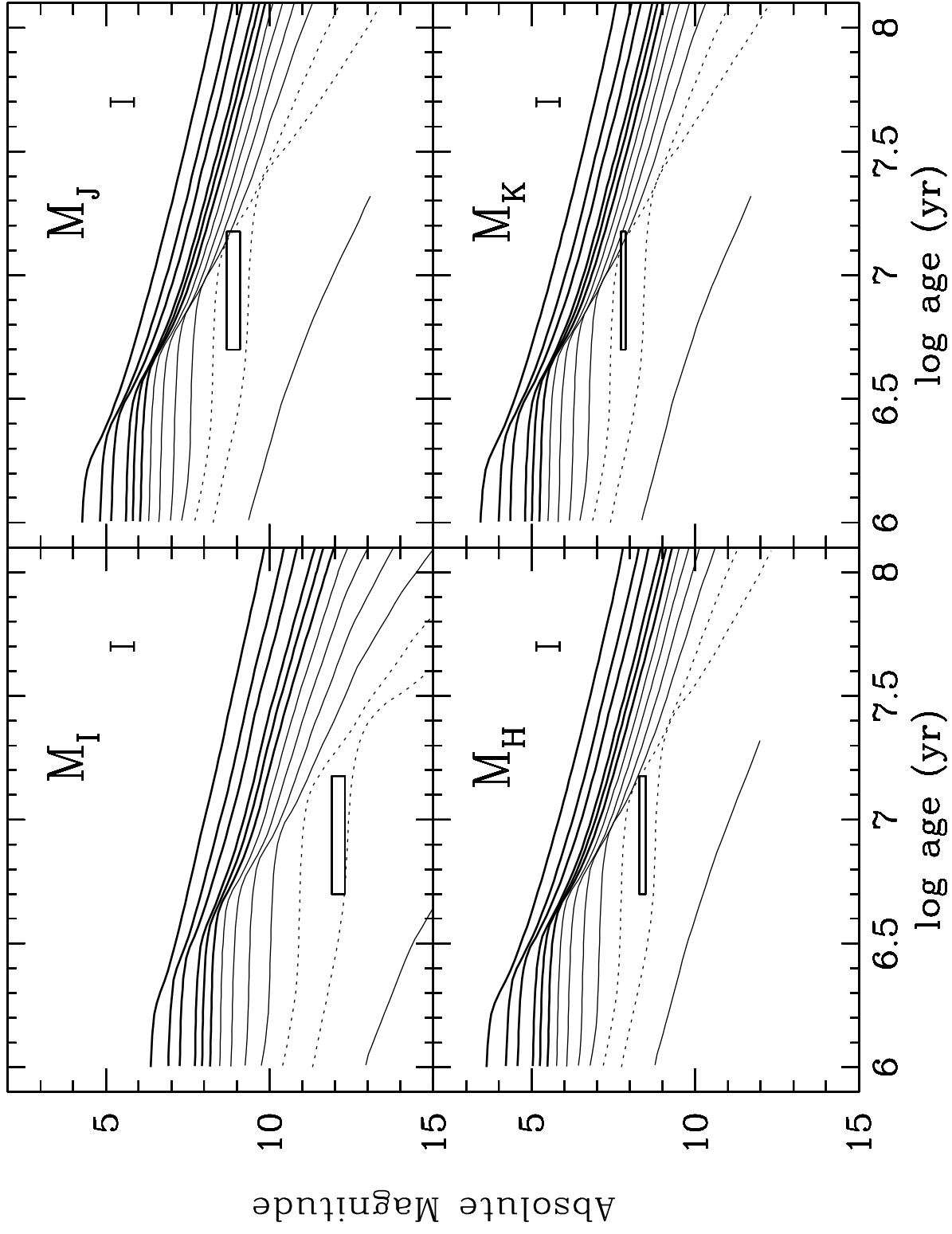
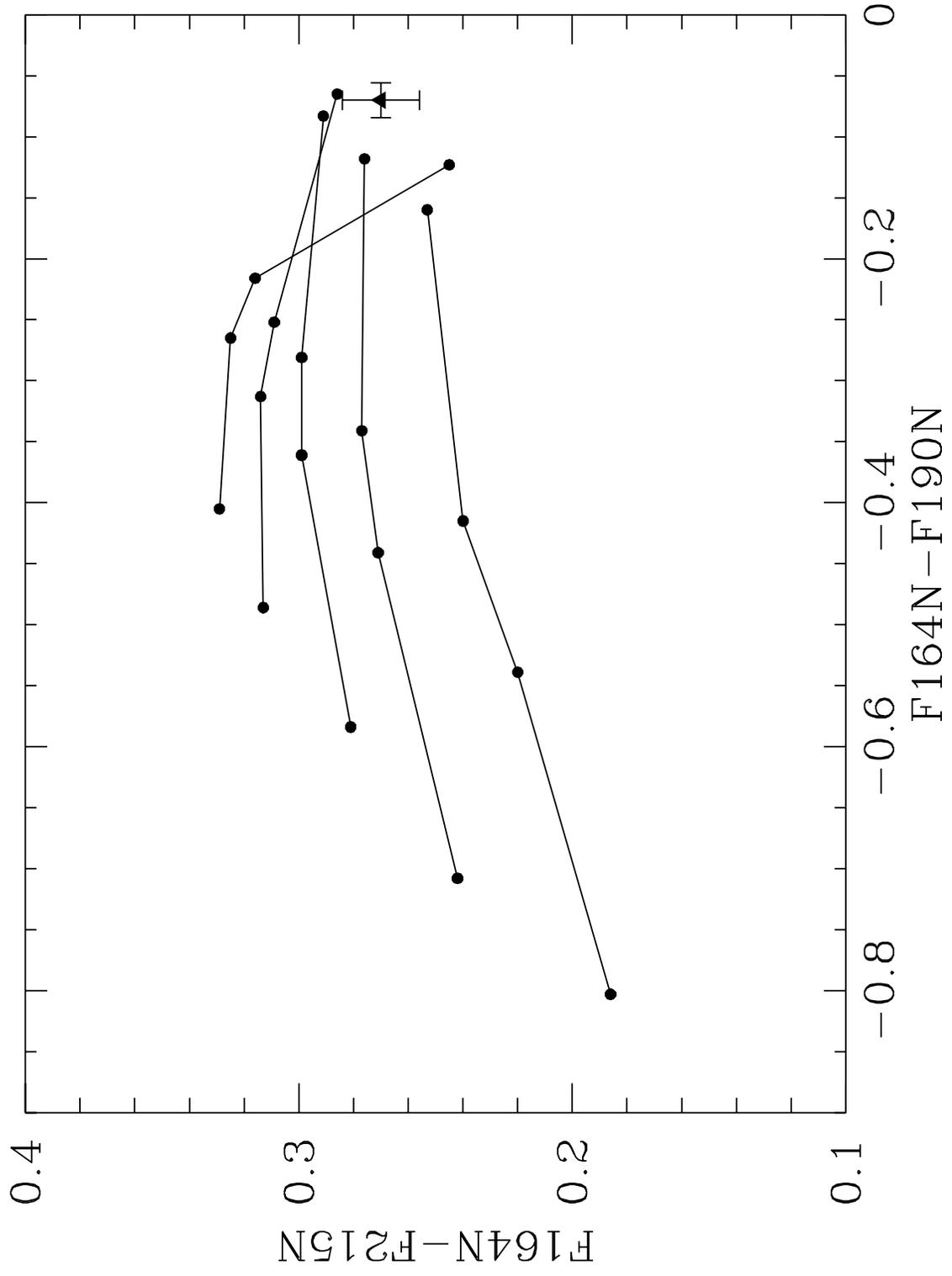
\end{document}